# Pump Linewidth Requirements for Processing Dispersion-Altered DQPSK Signals using FWM

SEÁN P. Ó DÚILL,* SEPIDEH T. NAIMI, LIAM P. BARRY

*The Rince Institute, Dublin City University, Glasnevin, Dublin 9, Ireland*
*\*Corresponding author: sean.oduill@dcu.ie*

We report on a potentially deleterious issue regarding the four-wave mixing (FWM)-based processing of dispersion-altered signals. We estimate the baudrate-dependent pump linewidth tolerances by calculating the extra optical signal to noise ratio (OSNR) penalty with respect to the propagation distance. We find that the issue is not important for 10 Gbaud differential quadrature phase shift keying (DQPSK) signals, though for 28 Gbaud (and 56 Gbaud) DQPSK signals we find that the pump linewidth requirements to implement FWM based optical signal processing needs to be in the sub-MHz range in order to avoid excessive OSNR penalties for the case of dispersion-altered signals. These results are pertinent for systems employing FWM, which could be all—optical wavelength converters for packet switching or mid-span spectral inversion techniques.

Optical signal processing using four-wave mixing (FWM) could play an important role within future high capacity optical communication systems [1]-[7]. A schematic of FWM is shown in Fig. 1(a) where a pump and signal beat together in a third-order nonlinear device resulting in the generation of an idler whose complex amplitude is the phase conjugate of the original signal. Wavelength conversion of signals with advanced modulation formats such as differential quadrature phase shift keying (DQPSK) [6],[7] and 16-state quadrature amplitude modulation (QAM) [1],[2] has already been achieved using FWM, and opens up the possibility of exploiting FWM to perform key network functions such as: (i) all-optical wavelength conversion in packet switched networks to avoid contention when packets of the same wavelength are being routed to the same output port; and (ii) mid-span spectral inversion (MSSI) [4],[5]. The concept of using an MSSI is shown in Fig. 1(b); whereby the FWM creates a complex conjugated copy of the signal that has traversed half-way along a fiber link so that the linear chromatic dispersion (CD) is compensated by transmission over the second half of the link, and if setup correctly, the fiber Kerr nonlinearity impairments could be cancelled [4]. MSSI is attracting renewed interest [4] to process signals in the optical domain to reduce the computationally-excessive digital signal processing required at the receiver to compensate for CD and nonlinear impairments, thereby reducing the latency of the data throughput.

In a realistic network, FWM may have to process a signal that has travelled many hundreds of kilometers, and the CD causes the pulses carrying the data symbols to spread into the time slots of neighboring symbols. The main impairment from FWM is the phase noise transfer from the pump to the idler; for degenerate FWM, the idler wave gets multiplied by double the phase noise on the pumps [6],[7]. In the scenario we consider here, the signal has been altered through CD with the original symbols spread over tens of symbol timeslots; the FWM

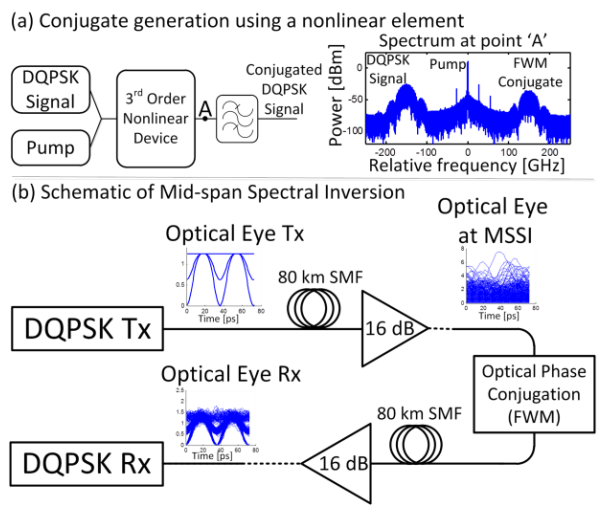

Fig. 1 (a) Typical technique to generate signal conjugation using FWM. The spectral placement of the signal and conjugate waves are shown relative to the pump. (b) Concept of the MSSI scheme, A clean DQPSK signal is generated and transmitted over many spans of SMF with inline amplification which adds noise to the signal. At the mid-span point, the accumulated fiber CD has distorted the waveform such that the combination of the phase conjugation and the second half of the fiber link undo the CD impairment as the signal reaches the receiver Rx. Simulated optical intensity eye diagrams are shown at the respective points along the link when the pump in the MSSI has zero linewidth.

process multiplies this dispersion-altered signal by the pump phase noise. This causes the received signal to be distorted because the original symbol cannot be fully recovered through the reversal of CD from either the fiber itself when using an MSSI, or through a digital CD compensator at the receiver. Most of the work so far on the FWM-based wavelength conversion of signals with advanced modulation formats has concerned signals that have not accumulated any CD [1],[2],[6],[7], and in those works the phase noise transfer simply adds to the existing phase noise that was on the signal and the signal can be demodulated provided that the transferred pump phase noise is within the range for the respective carrier phase recovery algorithms.

In this letter we estimate, through numerical simulations, the pump linewidth tolerance of the extra OSNR penalty for the FWM-based wavelength conversion of 10, 28 and 56 Gbaud DQPSK signals after transmission distances of up to 1,500 km over standard single mode fiber (SMF).. The extra OSNR penalty is calculated in detail for pumps with 1 and 2 MHz linewidth. Normally pump linewidths of a few megahertz result in negligible penalties for DQPSK signals in the absence of CD [7]; though the pulse spreading due to CD means that larger system penalties will now be incurred because the overlapped pulses are now multiplied by the phase noise and not completely equalized at the receiver nor through the second half of the fiber transmission span when using an MSSI. In order to reduce this effect, the coherence time of the laser phase noise (proportional to the inverse of the linewidth) must be very much longer than the duration of the number of pulses that overlap. This issue is problematic for large baudrate signals and we show that signals with larger baudrates, e.g. 28 and 56 Gbaud, incur severe system penalties. A pump linewidth of 2 MHz could incur an extra 6 dB OSNR penalty for 28 Gbaud DQPSK if the FWM element was placed at the mid-span length of 1,500 km. Even though we emphasize the results for MSSI subsystems, the results can be interpreted for FWM-based wavelength conversion in packet switched networks by noting that the mid-span distances could represent the distance between the switching nodes in the network.

A schematic of the simulation platform is shown in Fig. 1(b). We generate our signal waveform with $2^{18}$ DQPSK symbols. The optical intensity eyediagram is shown in the inset closest to the DQPSK transmitter showing the characteristic three levels for DQPSK at the symbol transition instances. The fiber is standard SMF with a loss of 0.2 dB/km and group velocity dispersion coefficient $\beta_2$ of 20 ps$^2$/km. The average launch signal power is set at 1 mW. The fiber propagation equation is given by

$$\frac{dA}{dz} = \frac{-\alpha}{2} A - j \frac{\beta_2}{2} \frac{d^2 A}{dT^2} \qquad (1)$$

Where $A$ is the slowly varying envelope of the optical field, $\alpha$ is the loss coefficient and $T$ is the time in a retarded time frame. The amplifier spacing is 80 km after which the signal is re-amplified by 16 dB using an ideal amplifier with a noise figure of 3 dB. The optical eye diagram of the dispersion-altered DQPSK signal is shown by the inset closest to the MSSI block in Fig. 1(b) indicating that the CD has completely altered the DQPSK signal. The optical field of the output of the optical phase conjugation element using FWM that includes the phase noise from the pump and ignores the constant frequency shift due to FWM is:

$$A_{mssi\_out}(T) = \eta_{FWM} A^*_{mssi\_in}(T) |E_P|^2 \exp\left(j 2\tilde{\phi}_p(T)\right) \qquad (2)$$

Where $A_{mssi\_in}$ is the input field to the mid-span spectral inverter and includes the effects of accumulated CD and additive amplifier noise, $\eta_{FWM}$ is the efficiency of the FWM process, $E_p$ is the field magnitude of the pump, and without loss of generality we assume that $\eta_{FWM} |E_P|^2 = 1$. The pump phase noise is given by $\tilde{\phi}_p$ and is assumed to be a Wiener-Levy process i.e. $\langle \tilde{\phi}_P(T+\tau) - \tilde{\phi}_P(T) \rangle^2 = 2\pi B_P |\tau|$ where $B_P$ is the linewidth of the pump. The essence of the problem associated with the phase noise transfer is explicit in Eq. (2). If the pump linewidth was zero, then no phase noise would be transferred and the propagation over the second half of the span would completely undo the signal alterations due to CD (as desired). When the pump linewidth is non-zero then the multiplicative nature of the phase noise transfer severely distorts the signal, the phase noise does not add to the original phase encoding of the DQPSK signal, and hence destroys the information encoded within. In Fig. 2 the signal distortion is highlighted by plotting the optical intensity eyediagrams at the receiver after the signal underwent FWM wavelength conversion at the mid-span lengths of 500, 1000 and 1500, km for the case of 28 Gbaud DQPSK signals. For the case of zero pump linewidth, the impairment is noise accumulation from the amplifiers;

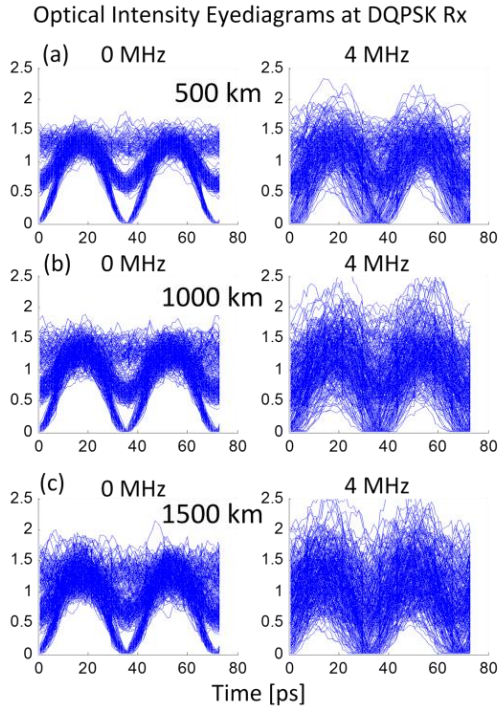

Fig. 2 Simulated optical intensity eyediagrams for the 28 Gbaud DQPSK signal at the receiver for a transmission system employing an MSSI. The mid-span lengths are 500, 1000 and 1,500 km. The diagrams on the left show the received eye intensity pattern when the pump linewidth in the MSSI is zero, the only impairment is additive noise from the amplifiers. On the right, the distorted eye patterns when using a pump with linewidth of only 4 MHz in the MSSI.

however when the pump linewidth is set to a mere 4 MHz are distorted eye diagrams are indicative of a loss of information.

The BER results for the received signal are calculated using a self-coherent receiver. To highlight the impact of converting dispersion-altered signals we first present the results for 28 GBaud because at this baudrate there are increasing though moderate penalties over mid span lengths ranging from 500 km to 1,500 km. The results are shown in Fig. 3. In Fig. 3(a) results for propagation over 0 km are shown and confirm that FWM using pumps with linewidths less than 4 MHz normally impose no system penalty on 28 Gbaud DQPSK signals. However, when the signal is wavelength converted after traversing mid-span lengths of 500 km (Fig. 3(b)), 1000 km (Fig. 3(c)) and 1500 km (Fig. 3(d)), then appreciable BER penalties accumulate. The general trend is that the BER increases significantly with increasing propagation distance and for increasing pump linewidth, especially for mid-span distances greater than 1,000 km. For the case of propagation over 500 km in Fig. 3(b), the 4 MHz pump causes a 2.5 dB penalty at the BER of $10^{-4}$, and the issue of converting dispersion-altered signals might not be an issue for exploiting FWM as wavelength conversion elements within packet switched networks. When propagating over distances of 1,000 km then clear BER floors emerge which are the likely transmission distances after which an MSSI would be employed; therefore care would be needed to select lasers with narrow linewidths to construct MSSI elements.

There will be baudrate dependent penalties due to the quadratic dependence of CD on the signal bandwidth. We repeat the simulations again at the lower baudrate of 10 Gbaud. Here there is less pulse spreading than for the 28 Gbaud case and we just concentrate on transmission over half-span lengths of 1,000 km and 1,500 km, the results are shown in Fig. 4. Even over propagation of 1,500 km, that only a moderate extra OSNR penalty of 1 dB is incurred for a pump linewidth of 4 MHz. Therefore the pump linewidth would not be a limiting factor in realizing

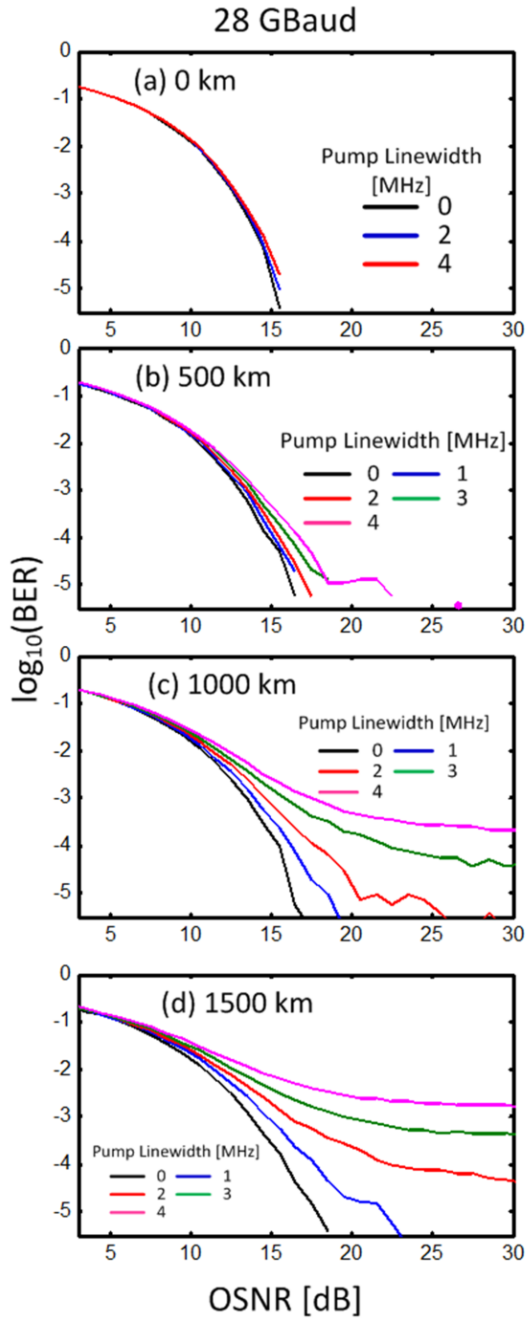

Fig. 3 Calculated BER results for noise-loading the DQPSK signal at the receiver for operating an MSSI with mid-span lengths of (a) 0 km, (b) 500 km, (c) 1,000 km and (d) 1,500 km. The pump linewidth in the MSSI is used as parameter. For case (a): these results show that a pump linewidth of 4 MHz does not cause any OSNR penalty when the signals have not accumulated any chromatic dispersion. (b) – (d) The BER increases with increasing mid-span length and with increasing pump linewidth. Large OSNR penalties occur especially for lengths > 1000 km, N.B. Appreciable OSNR penalties are experienced even for using pumps with linewidths of 2 MHz.

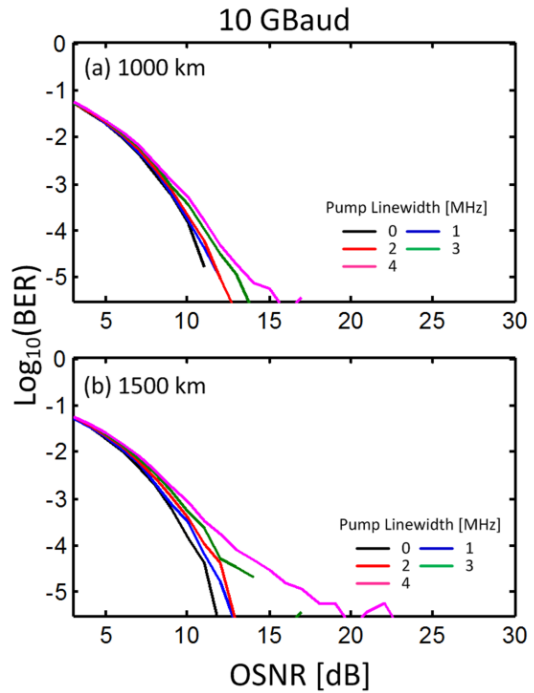

Fig. 4 Calculated BER results for noise-loading a 10 Gbaud DQPSK signal at the receiver for operating and MSSI with mid-span lengths of: (a) 1000 km and (b) 1,500 km. The pump linewidth in the MSSI is used as parameter. These results show that a modest OSNR penalty is incurred only when a pump with a linewidth of 4 MHz is used and that the DQPSK signal has travelled 1,500 km.

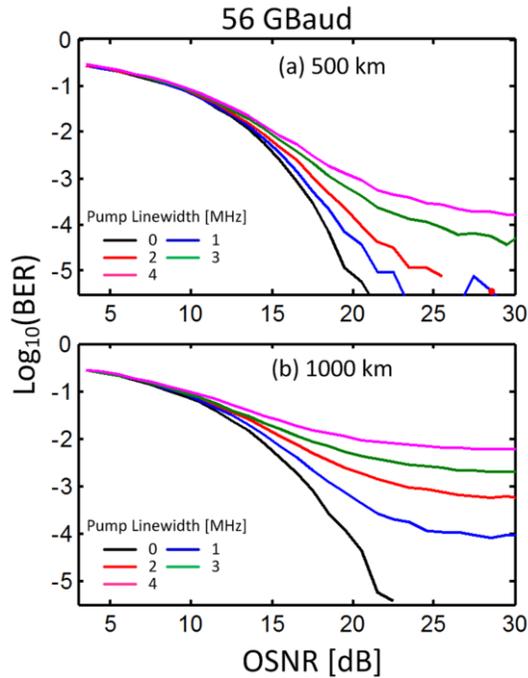

Fig. 5 Calculated BER results for noise-loading a 56 Gbaud DQPSK signal at the receiver for operating an MSSI with mid-span lengths of: (a) 500 km and (b) 1,000 km. The pump linewidth in the MSSI is used as parameter. These results show that unlike the case for 10 Gbaud and 28 Gbaud systems that large OSNR penalties are incurred even when even for mid-span lengths of 500 km. For mid-span distances of 1,000 km, that BER floors are observed for pump linewidths of 1 MHz.

MSSI elements for 10 Gbaud DQPSK signals. However the situation is very much different for 56 Gbaud DQPSK signals. The increased signal bandwidth causes greater pulse spreading and therefore we expect larger transmission penalties. The results are shown in Fig. 5 for half-span propagation distances of 500 km and 1,000 km. BER floors are clearly visible even for mid-span propagation distances of 500 km, which indicate that care will be needed to carefully select pumps with linewidths below 3 MHz for FWM-based wavelength conversion subsystems for packet switched networks. For mid-span lengths of 1,000 km, clear error floors emerge above the BER of $10^{-4}$ even for the case of 1 MHz pumps, which would dissuade the use of MSSI elements to equalize 56 Gbaud signals.

We summarize the findings by calculating the mid-span distance dependence of the extra OSNR penalty at the BER of $10^{-4}$ for the pump linewidths of 1 and 2 MHz. The extra OSNR penalty is the difference in dB between the OSNR that yields a BER of $10^{-4}$ compared to the required OSNR for pumps with zero linewidth to achieve the same BER of $10^{-4}$. The results are shown in Fig. 6. For 10 Gbaud DQPSK signals (black curves), the maximum extra OSNR penalty is 1 dB for the mid-span length of 1,500 km when using a pump with 2 MHz linewidth; therefore FWM would be suitable to process dispersion-altered 10 Gbaud DQPSK signals irrespective of the length of the fiber span. For 28 Gbaud DQPSK signals (blue curves) the OSNR penalties rise sharply for mid-span lengths exceeding 1,000 km and when using a 2 MHz pump. Therefore pumps with sub-MHz linewidths would be needed to construct MSSI elements to process 28 Gbaud DQPSK signals. Nonetheless, over mid-span propagation distances ~500 km, only an extra 0.5 dB penalty is incurred when using a pump with 2 MHz linewidth, therefore FWM-based wavelength conversion for packet switched networks of 28 Gbaud DQPSK signals would be feasible. Unfortunately for 56 Gbaud DQPSK signals (red curves), excessive extra OSNR penalties >5 dB are incurred for the longer mid-span lengths of 1000 km when using pumps with 1 MHz linewidth. Over the mid-span length of 500 km, that even the 1 MHz linewidth pumps incur an extra OSNR penalty of 0.5 dB. Extra simulations for 56 Gbaud DQPSK systems show that pumps with 250 kHz linewidth would incur extra OSNR penalties of 0.2 dB and 1 dB at the mid-span lengths of 500 km and 1,500 km respectively.

We calculated the extra OSNR penalties associated with the wavelength conversion of dispersion-altered DQPSK signals. While the issue is not serious for 10 Gbaud DQPSK systems, special attention needs to be placed on selecting pumps with sub-MHz linewidths for FWM-based wavelength converters when operating on 28 Gbaud signals that have traversed 1,000 km and on 56 Gbaud DQPSK signals that have traversed 500 km.

**Funding.** Science Foundation Ireland (SFI). (12/RC/2276)

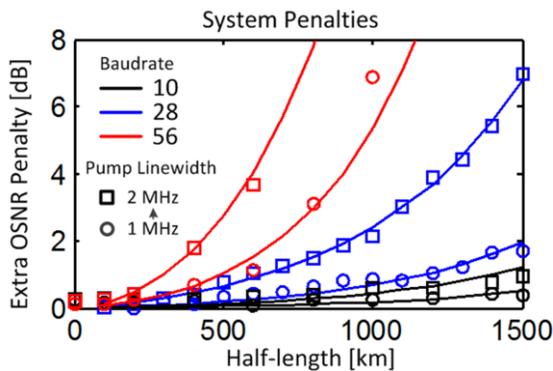

Fig. 6. Summary of the calculated excess OSNR penalty at the BER of $10^{-4}$ for mid-span lengths ranging from 100 to 1500 km. The penalties are calculated for pump linewidths of 1 and 2 MHz for DQPSK baudrates of 10, 28 and 56 Gbaud. Symbols are simulation results, the lines are guides.